\documentclass[a4paper,10pt]{article}

\usepackage{helvet}         
\usepackage{courier}        
\usepackage{type1cm}        
\usepackage{makeidx}         
\usepackage{graphicx}        
\usepackage{float}
\usepackage{multicol}        
\usepackage[bottom]{footmisc}
\usepackage[utf8]{inputenc}
\usepackage[T1]{fontenc}
\usepackage{amsmath}
\usepackage{amssymb}
\usepackage{graphicx}
\usepackage{hyperref}
\usepackage{geometry}
\usepackage{authblk}  
\usepackage{amsmath}
\usepackage{algpseudocode} 
\usepackage{algorithm}
\usepackage{bm}
\usepackage{amssymb}
\usepackage{rotating}
\usepackage{multirow}
\usepackage{xcolor}
\usepackage{placeins}
\usepackage{natbib}
\usepackage{subcaption}
\newtheorem{defn}{Definition} 

\begin{document}

\title{\textbf{Random Survival Forest for Censored Functional Data}}

\author[1]{Elvira Romano}
\author[2]{Giuseppe Loffredo}
\author[3]{Fabrizio Maturo}

\affil[1]{Department of Mathematics and Physics, University of Campania "Luigi Vanvitelli", Caserta, Italy}
\affil[2]{Department of Mathematics and Physics, University of Campania "Luigi Vanvitelli", Caserta, Italy}
\affil[3]{Department of Economics, Statistics and Business, Faculty of Technological and Innovation Sciences, Universitas Mercatorum, Rome, Italy}

\date{}

\maketitle

\begin{abstract}
This paper introduces a Random Survival Forest (RSF) method for functional data. The focus is specifically on defining a new functional data structure, the Censored Functional Data (CFD), for dealing with temporal observations that are censored due to study limitations or incomplete data collection. This approach allows for precise modelling of functional survival trajectories, leading to improved interpretation and prediction of survival dynamics across different groups. A medical survival study on the benchmark SOFA data set is presented. Results show good performance of the proposed approach, particularly in ranking the importance of predicting variables, as captured through dynamic changes in SOFA scores and patient mortality rates.
\end{abstract}


\footnotetext{\textbf{Abbreviations:} FDA, Functional Data Analysis; FPC, Functional Principal Components; FCTs, Functional Classification Trees; FRF, Functional Random Forest; FBG, Functional Bagging; OOBFD, Out of Bag Functional Data; IBFD, In-Bag Functional Data; ESC, Empirical Splitting Curve; TSC, Theoretical Splitting Curve; FBGSS, Functional Between Groups Sum of Squares; FBLSS, Functional Between Leaves Sum of Squares.}

\section{Introduction}
\label{sec1}

Survival analysis focuses on estimating and predicting the time until an event of interest, such as death or another unique occurrence, using historical data \cite{Kleinbaum2012}. This type of analysis can be challenging because the exact timing of the event may be unknown in certain instances.  Survival trees (STs), a specific type of model within this field, allow for uncovering intricate nonlinear relationships in an intuitive, interpretable format. By iteratively dividing the population into subgroups and predicting a unique survival distribution for each terminal node, STs provide a powerful tool for understanding and forecasting time-to-event data.
STs and their ensembles are widespread in non-parametric modelling. STs are particularly noteworthy for their ability to handle complex relationships and interactions among variables \cite{Ishwaran2008}. They offer a clear picture of the decision-making procedure, making it easier to interpret survival patterns and recognise critical prognostic factors. As a result, STs have established a unique position in survival analysis by delivering a visually understandable framework for modelling and predicting survival probabilities. 
Well-known limitations of single STs are variance and overfitting. To address these concerns, Bagging (SB) and Random Survival Forest (RSF) \cite{Wang2019} are frequently used. These techniques enhance reliability by averaging multiple trees, reducing variance, preventing overfitting, and improving prediction accuracy and robustness. An extensive review can be found in \cite{Wang2017}. SB trains multiple STs on bootstrapped samples generated from the original dataset. Then, the single predictions of STs are combined through averaging or voting.
RSF expands the concept of traditional Random Forest (RF) to survival analysis \cite{Isra}, \cite{Biemann2010}. In RSF, each tree is trained on a random subset of observations and a random subset of features at each split. Introducing randomness in the tree-building process, RSF diminishes the correlation between STs and improves the diversity of the ensemble, resulting in better generalisation performance. Further research also focused on combining survival analysis with various modelling approaches, discussed in more detail in \cite{10.1093/biostatistics/kxv001}.

Studying the relationship between time-varying processes and time-to-event data is challenging in clinical research. A modern alternative solution involves using Functional Data Analysis (FDA) \cite{Ramsay2005}. 
FDA transforms longitudinal data into curves,  which are then integrated into survival analysis models. A first approach proposed by \cite{Lin2021} incorporates multiple longitudinal outcomes by extracting features using two methods: multivariate functional principal component analysis and multivariate fast covariance estimation for sparse functional data. These extracted features are subsequently used as covariates in a survival model, with FPCA aiding in the estimation process. 
Spreafico et al. (2023) \cite{Spreafico2023} proposed a novel approach for addressing time-varying covariates in survival analysis using FDA.  The latter research aims to transform longitudinal data into functional data, facilitating their treatment and analysis within the survival framework. 
For longitudinal studies, patients often return for follow-up visits at regular intervals. Therefore, the true event time is only known to lie within an interval between visits, and the exact time is obscured. Multiple studies have demonstrated biased survival outcome estimates using the Cox model's well-developed methodology for modelling interval-censored data \cite{zhang2010interval}. In many studies, the focus has been on reconstructing curves over the entire follow-up period of study \cite{delaigle2013classification}\cite{strzalkowska2021censored} to manage the data at the same observation time.
Starting from these latter ideas, our research strives to extend the concept of RSF to the functional data framework, exploiting the FDA's potential in dimensionality reduction, predictive power, interpretability, and ability to reconstruct censored data.

The paper positions itself within the literature on combining FDA and statistical learning. The starting point of our proposal is to extend the Functional Random Forest (FRF) offered by \cite{Maturo2023},   \cite{MaturoVerde} to the context of irregular data in survival analysis,  with a particular focus on the issues of censured data reconstruction and explainability of the RSF. 

Implementing RSF in the functional context poses significant challenges, especially given the vast amount of available data and the incomplete nature of each statistical unit's temporal sequences.   
This work considers a new type of functional data, defined as Censored Functional Data (CFD).  Unlike previous studies, we concentrate on the actual observation period to reconstruct the trajectories associated with each unit rather than extending the curves over the entire temporal domain of the follow-up study. This approach aims to maximise the utilisation of available information during the study period without excessive extensions or interpolations.
Essentially, the main goal is to accurately model the trajectory of curves, considering only the observation period associated with each unit, thus ensuring a precise representation of the survival dynamic within that specific period.

By integrating the CFD with FPCA, we aim to capture relationships and dynamic patterns in the data, leading to more accurate and interpretable survival predictions within STs. Section 2 introduces preliminaries on Functional Data Analysis and Survival Random Forest and describes our contribution. Section 3 assesses the performance of our proposed approach on the well-known "SOFA" data set. The paper ends with a discussion and conclusion on our proposal.

\section{Material and Methods}
\label{FDAsec}

\subsection{Preliminaries on Functional Data Analysis (FDA)}

The field of Functional Data Analysis (FDA) \cite{Ramsay2009,Ramsay2005,Ferraty2011}  deals with data represented in functional form rather than using traditional discrete observations. Typically, data are handled as vectors or matrices, with an assumption of independence among observations. However, in many situations, the observations are displayed on a discrete time scale or spatial domain, and the information is not available on the entire study domain. Indeed, FDA aims to capture the underlying structure and the variability present in discrete data by transforming it into a functional data object that assumes values within a functional space denoted as $\mathcal{H}$. In general, the functional data object can be represented as a random variable $ \mathcal{X} = \{ X(t); t \in \mathcal{T} \} $, where $\mathcal{T}$ is a common domain and $X(t)$ denotes the value that the function assumes at time or position $t$. In this article, we will consider $\mathcal{T}$ to be a space-time domain. This real-valued function can be viewed as the realization of a one-dimensional stochastic process, often assumed to be in a Hilbert space, such as $L^2$ defined on the compact domain $\mathcal{T}$. Here, a stochastic process $\mathcal{X}$ is said to be an $L^2$ process if and only if it satisfies $E\big[\int_{\mathcal{T}} X(t)\, dt\big] < \infty$. The $L^2$ space is defined as a normed space of functions where the norm is given by $\|\cdot \|_2$. In general, considering a time domain $\mathcal{T}$, in the Hilbert space $(L^2(\mathcal{T}),\|\cdot \|_2)$ is defined the distance between two functions $f(t)$ and $g(t)$  as follows:
\begin{equation}
d_{L^2}(f,g) =\| f-g \|_{2}= \left( \int_{\mathcal{T}} \left| f(t) - g(t) \right|^2 dt \right)^{\frac{1}{2}}\,.
\end{equation}

FDA aims to convert discrete data into a functional form using techniques such as smoothing, interpolation, and regression \cite{Schimek2013}. We observe the discrete-time data given by $N$ pairs $(Y_i,t_i)$, where $t_i\in \mathcal{T}$ and $Y_i\in \mathbb{R}$ are the values recorded at the time $t_i$ for $i=1,...,N$ statistic units. The $i$-th recorded value $Y_i=Y(t_i)$ corresponds to a realization of a generic function $X(\cdot)$ at time $t_i\in\mathcal{T}$. In FDA, we can define the function  $X(\cdot)$ by finite linear combination \cite{Ramsay2005} as follows:
\begin{equation}
X(t) = \sum_{j=1}^{K} c_j \phi_j(t)\quad t\in \mathcal{T} 
\label{sum_bas}
\end{equation}
where $\{ \phi_j(t) \}_{j=1}^{K}$ is a set of basis functions used in the representation, and $\{ c_j \}_{j=1}^{K} $ are the coefficients of each basis function. However, in real cases, the $i$-th observation can be affected by the $i$-th term's error, $\varepsilon_i$, due to measurement inaccuracies or other factors introducing variability. Thus, we observe $Y_i$ as $Y_i=X(t_i)+\varepsilon_i$.
In the Equation (\ref{sum_bas}), the functional data is approximated by a finite linear combination of basis functions that can be chosen based on the specific problem and the characteristics of the data trend. The coefficients $\{ c_j \}_{j=1}^{K}$ can be calculated using the sum of squared errors (SSE), where the problem is given by:
\begin{equation}
\min\, (\varepsilon_i)^2=\min_{c_1, c_2, \ldots, c_K} \sum_{i=1}^{N} \left( Y_i - \sum_{j=1}^{K} c_j \phi_j(t_i) \right)^2\,,
\end{equation}
where $N$ is the number of data points observed, $t_i$ represents the $i$-th instant time, and $K$ is the number of basis functions chosen for the functional representation. In the context of FDA, considering the $N$ data functions expressed as $X_1(t),..., X_N(t)$, we can introduce the standard estimators for the mean and covariance functions:
\begin{equation}
\label{2}
\hat{\mu}(t)=\frac{1}{N}\sum_{i=1}^N X_i(t)
\end{equation}
and
\begin{equation}
\widehat{C}(t,s)=\frac{1}{N}\sum_{i=1}^N [(X_i(t)- \hat{\mu}(t))(X_i(s)- \hat{\mu}(s))]\,.
\end{equation}\label{3}
Moreover, the FDA provides many useful tools for dealing with functional data to interpret and make predictions e.g., the Functional Principal Component Analysis (FPCA) \cite{Ramsay2005, Ferraty2006}, Functional Regression Analysis (FRA) \cite{Ramsay2005, Febrero2012} and Functional Classification Trees (FCTs) \cite{ Maturo2023, MaturoVerde, maturo2024combining}. FPCA extends classical Principal Component Analysis (PCA) in a functional context. The method allows us to reduce the high dimensionality of the data, preserving the maximum amount of information \cite{Ramsay2005, Aguilera2013, Febrero-Bande2012}. In this context, the approximated functional data $X(\cdot)$ is given by:
\begin{equation}
\label{2bis}
X(t) = \sum_{m=1}^{p} \nu_{m} \xi_m(t)\quad t\in \mathcal{T}\,,
\end{equation}
where $p$ represents the total number of Functional Principal Components Scores (FPCs), where $\xi_{m}(t)$ denotes the eigenfunction for the $m$-th of the function $X(t)$.  The explained variance of these curves is determined by $\sum_{m=1}^{p} \lambda_{m}$, where $\lambda_{m}$ represents the variance of the $m$-th FPC. It's important to note that the variance explained by each FPC decreases as the index $m$ increases, that is $\lambda_1\geq\lambda_2\geq\ldots\geq\lambda_p$. We assume $X_i(t)$ functions, for $i=1,\ldots,N$, be centered, i.e., $\int_{\mathcal{T}} X_i(t) dt = 0$, which implies that the overall mean function $\mu(t)$ is zero. The $m$-th FPC can be obtained as follows:
\begin{equation}
\nu_{im} = \int_{\mathcal{T}} X_i(t) \xi_m(t) dt, \quad i = 1, \ldots, N, \quad m= 1, \ldots, p 
\end{equation}
where the eigenfunctions $\xi_m(t)$ are obtained resolving this problem: 
\begin{equation}
\max_{\xi_{m}} Var \Bigg[ \int_{\mathcal{T}} X_i(t) \xi_m(t) dt \Bigg],
\end{equation}
s.t.
\begin{equation}
\|\xi_{m}\|_2^2=\int_\mathcal{T} \xi_{m}(t)^2\,dt =1\, ,
\end{equation}
and
\begin{equation}
 \int_\mathcal{T} \xi_{m}(t)\xi_{n}(t)\,dt=0\,\qquad \text{for}\qquad m\neq n.
\end{equation}

In statistics and machine learning, the data points can be classified into distinct categories using the features collected during the study, which can vary across different fields and applications. In many situations, utilizing functional data objects as features can be advantageous. Starting from a classic Classification Tree (CT), we can move to the functional approach, which involves adapting traditional techniques to accommodate functional data, which are functions defined over a continuous domain. The method is called Functional Classification Tree (FCT) and allows us to deal with functions defined in a functional space $L^2(\mathcal{T})$. The methodology takes the pairs $\{Z_i, X_i(t)\}$, where $X_i(t)$ is a predictor curve for $i$-th unit defined within metric space $(L^2(\mathcal{T}),\Vert\cdot\Vert)$, and $Z_i$ is a scalar response observed for $i$-th unit. Thus, the FCT algorithm predicts the vector-response $\bm{Z}\in\mathbb{R}^N$ considering as features the FPCs' scores obtained from decomposition (\ref{2bis}) applied for each $X_i(t)$ to obtain the score-matrix as follows:

\begin{equation}
\bm{V}=
\begin{pmatrix}
\nu_{11} & \nu_{12} & \dots & \nu_{1p}\\
\nu_{21} & \nu_{22} & \dots & \nu_{2p}\\
\vdots & \vdots & \ddots & \vdots \\
\nu_{N1} & \nu_{N2} & \dots & \nu_{Np}\\
\end{pmatrix}
\end{equation}

\noindent where $\bm{V}\in \mathbb{R}^{N\times p}$ and $\nu_{im}$ correspond to FPC score of the $i$-th curve relative to the $m$-th eigenfuncion $\xi_m$ for $i=1,...,N$ and $m=1,...,p$ \cite{MaturoVerde, Maturo2023}. This approach is called the Functional Classification Tree with Principal Components (FCT-FPCs). The FCT-FPCs organize data into groups based on their characteristics. Starting with all data points, which are the functions, the algorithm involves stratifying and segmenting the predictor space into rectangular regions, called terminal nodes or lives, based on certain criteria that make the groups more homogeneous by using as features the score-vector $\bm{\nu}_m=(\nu_{1m},...,\nu_{Nm})^T$ for $m=1,...,p$. 
The method selects the best features $\bm{\nu_m}$ and the best threshold $c$ to define at each internal node of the tree two half-planes given by:
\begin{equation}
R_1(m,c)=\{X(t)\in L^2(\mathcal{T}) | \bm{\nu_m}\leq c\}\quad\text{and}\quad R_2(m,c)=\{X(t)\in L^2(\mathcal{T}) | \bm{\nu_m} > c\}
\end{equation}
Then, to determine the best feature and threshold to split on, the algorithm uses the splitting criterion such as the Gini index or the Shannon-Weiner index \cite{chao2003nonparametric, Chao2014}. Let $Q(\text{node})$ be the splitting criterion at node $h$, which divides it into two daughter nodes $\text{left-daughter}$, $\text{right-daughter}$. Let $Q(\text{left-daughter})$, $Q(\text{right-daughter})$ be the splitting criteria for the node and its two daughter nodes. The impurity is computed as follows:
\begin{equation}
\Delta Q = Q(\text{node}) - \left( \frac{n_{\text{left}}}{n_{\text{node}}} Q(\text{left-daughter}) + \frac{n_{\text{right}}}{n_{\text{node}}} Q(\text{right-daughter}) \right)
\end{equation}

Where $n_{\text{node}}$ is the number of samples in the splitting-node, $n_{\text{left}}$ and $n_{\text{right}}$ are the number of samples in the left and right daughter nodes. The feature $\bm{\nu}_m$ and threshold $c$ that maximize $\Delta Q$ are chosen for splitting at each node.

\subsection{Contribute}
This section focuses on managing survival time data, starting with the creation of censored functional data and ending with the creation of the Functional Random Survival Forest (FRSF) based on FPCs. We explore the complexity of managing survival data, address challenges posed by censoring, and use techniques such as Principal Components Analysis through Conditional Expectation (PACE) to extract meaningful features from irregular functional data. 


\subsubsection{Censored Functional Data}
\label{CFD}
In a survival study, $N$ subjects are enrolled in a given period  $\mathcal{T}=[a,b]$ called a follow-up. In this period, different data are collected for each $i$-th subject, with $i=1,\ldots,N$ . In general, the times $T_i$ denote the survival time, and $C_i$ denote the censored time, i.e., an event did not occur for the subject. Let $\delta_i$ be defined for each subject as follows:
\begin{equation}
\label{1}
\delta_i:=\begin{cases}
1 & \text{if $T_i\leq C_i$ (Event occurred)}\\
0 & \text{if $T_i>C_i$ (Censoring occurred)}
\end{cases}
\quad i=1,...,N\,.
\end{equation}

During the follow-up study, $J_i$ values are collected for each subject until the event time, denoted by $T_i^*=\mbox{min}\,(C_i,T_i)$, occurs. This information defines the Survival Time Data (STD) given by $(\bm{Y}_{i},\bm{t}_i,T_i^*,\delta_i)$, where $\bm{Y}_{i}=(Y_{11},...,Y_{1J_i})^T$ refers to the observed values vector with $Y_{ij}=Y(t_{ij})$ and $\bm{t}_{i}=(t_{i1},...,t_{iJ_i})^T$ is the time vector for the $i$-th unit. 

In this context, the main challenge of employing the FDA approach and leveraging functional tools is to retrieve data for every subject, even when there are no observations during the entire observation period. For this reason, our approach involves the introduction of CFD, enabling the continuous reconstruction of information for each subject at any given moment. The main issue of this extension is how to replace these data by using different functions according to the number of recording values $J_i$.
Note that $t_{iJ_i}\leq T_i^*\leq t_{iJ_{i+1}}$ for each $i$-th unit and no longitudinal measurements are available after $T_i^*$. Thus, the observation of the function $\mathcal{X}_i$ consist of $J_i$ pairs $(t_{ij},Y_{ij})$ for $i=1,...,N$ and $j=1,\ldots,J_i$. 

Constructing functional data requires a different approach in survival analysis with censored data. Hence, we aim to develop a methodology that handles censoring and addresses the unique aspects of functional data, such as temporal truncation, to ensure accurate modelling and interpretation of the underlying patterns over time. In our case, we have $J_i\neq J$ as observed data, which are $J_i$-dimensional. For this reason, the first step is to convert the observed values $\{(Y_{ij},t_{ij}): i=1,..., N\: \text{and}\: j=1,..., J_i\}$ into a functional form by considering different recording values. In this connection, the standard approach to estimating the functional form by starting with observed values is the basis approximation. The CFD can be defined as follows:

\begin{defn}
\label{def1}
Let functional space $L^2(\mathcal{T})$ where $\mathcal{T}=[a,b]$ is a compact interval. The CFD is a functional data expressed as follows:
\begin{equation}
\label{4}
\mathcal{X}_i=\{X_i(t): t\in [a,T_i^*]\}\quad \text{with}\quad\:T_i^*=\mbox{min}\,(C_i,T_i)\quad \text{and} \quad\delta_i=1_{[T_i^*=T_i]}\quad \text{for}\quad i=1,...,N
\end{equation}
where $T_i$ is the true event time and $C_i$ is the censored time.
\end{defn}

As in (\ref{4}), we can express these functions according to numbers of $J_i$. For which, using a fixed basis representation and other functional representations, we obtain the CFD as follows:

\begin{defn}
\label{defn2}
Let functional space $L^2(\mathcal{T})$ defined on the compact interval $\mathcal{T}=[a,b]$ and pairs observed values $(t_{ij},Y_{ij})$, where $t_{ij}\in \mathcal{T}$ for $j=1,..., J_i$, the CFD is built as follows:
\begin{equation}
\label{5}
X_i(t):=\begin{cases}
Y_{ij} & \text{if $J_i=1$}\\
\beta_0+\beta_1 t & \text{if $J_i=2$}\\
\sum_{k=1}^{K_i} c_{ik}\phi_k(t) & \text{if $J_i\geq 3$}
\end{cases}
\quad\text{with}\quad t\in[a,T_i^*]\quad \text{and} \quad\delta_i=1_{[T_i^*=T_i]}\quad\text{for}\quad i=1,...,N\
\end{equation}
where $c_{ik}$ corresponds to the $k$-th coefficient for $i$-th unit and $\phi_k(t)$ to $k$-th basis function that can be chosen to approximate the full basis expansion. 
\end{defn}
Moreover, $K_i$ is determined individually for $i$-th unit when the number of recorded values exceeds or equals $3$. It is determined using leave-one-out cross-validation. The method minimizes the prediction error of the number of components $K$ choosing the best $\hat{K}_i$ for the $i$-th subject as follows: 

\begin{equation}
    \hat{K}_i = \arg \min_{K} \sum_{j=1}^{J_i} \left\{ Y_{ij} - \hat{Y}_i^{(-i)}(t_{ij}) \right\}^2
    \quad\text{for}\quad i=1,\ldots,N
\end{equation}

In addition, considering that in real-world applications the measurement of $Y_{ij}$  may be subject to errors, we consider the random noise $\varepsilon_{ij}$ with $E[\varepsilon_{ij}]=0$ and $var[\varepsilon_{ij}]=\sigma_{ij}^2$, where $\varepsilon_{ij}$ are independent across $i$ and $j$ and the errors are considered homoschedastic with $\sigma_{ij}^2=\sigma^2$. Then, we can express the observed values as $Y_{ij}=X_i(t_{ij})+\varepsilon_{ij}$, with $t_{ij}\in [a,T_i^*]$, max$\{T_i^*:i=1,...,N\}\leq \tau$, where $\tau$ is the length of the study follow-up. Hence, for different numbers $J_i\geq 2$ using the definition (\ref{5}) we obtain the follow model:
\begin{equation}
\label{6}
Y_{ij}=\begin{cases}
\beta_0+\beta_1 t_{ij}+\varepsilon_{ij} & \text{if $J_i=2$}\\
\sum_{k=1}^{K_i} c_{ik}\phi_k(t_{ij})+\varepsilon_{ij} & \text{if $J_i\geq 3$}
\end{cases}
\quad\text{with}\quad t_{ij}\in[a,T_i^*]\quad \text{for} \quad i=1,...,N\,\text{,}\,j=1,...,J_i\,.
\end{equation}
In order to evaluate the coefficients for the (\ref{6}), we use the SSD criterion as follows:
\begin{equation}
\min \,\Bigg[\sum_{j=1}^{J_i}[Y_{ij}-X_i(t_{ij})]^2\Bigg]=\begin{cases}
\min_{\beta_0,\beta_1}\,\big[\sum_{j=1}^{J_i}[Y_{ij}-\beta_0-\beta_1 t_{ij}]\big] & \text{if $J_i=2$}\\
\min_{c_{i1},c_{i2},...,c_{iK_i}}\,\big[\sum_{j=1}^{J_i}[Y_{ij}-\sum_{k=1}^{K_i} c_{ik}\phi_k(t_{ij})]\big]& \text{if $J_i\geq 3$}
\end{cases}
\end{equation}
calculated for all subjects  $i=1,...,N$.

\subsubsection{Principal component analysis for Censored Functional Data}

FPCA is an extension of the classical PCA because the principle is to replace vectors with functions, matrices by linear operators, and, in particular, covariance by auto-covariance operators \cite{Ramsay2005}. Moreover, scalar products in vector are replaced by scalar products in function space $L^2$. Let's suppose that  $X(t)$ is a generic trajectory defined on Hilbert space $L^2(\mathcal{T})$ with its mean function $\mu(t)=E[X(t)]$ and its covariance function $G(s,t)=Cov(X(t), X(s))$, with $t,s \in \mathcal{T}$. The covariance function can be expressed by the spectral decomposition as $G(s,t)=\sum_{m=1}^\infty\lambda_m\xi_m(t)\xi_m(s)$, where $\{\lambda_m\}_{m=1,..,\infty}$,  correspond to a set non-increasing eigenvalues such that $\sum_{m}\lambda_m <\infty$ and $\{\xi_m(t)\}_{m=1,..,\infty}$ correspond to the eigenfunctions. Thus, the trajectory $X(t)$ admits the following Karhunen-Loevè expansion:
\begin{equation}
\label{7}
X(t)=\mu(t)+\sum_{m=1}^\infty\nu_{m}\xi_m(t)\quad \text{with}\quad t\in \mathcal{T}\,,
\end{equation}
where the coefficient $\nu_{m}=\int_{\mathcal{T}}[X(t)-\mu(t)]\xi_m(t)\,dt$  corresponds to the $m$-th Functional Principal Component (FPC) score of $X(\cdot)$ and satisfies the condition $E[\nu_{m},\nu_{n}]=\delta_{m\,n} n_m$ with $\delta_{m\,n}=1$ if $m=n$ and $0$ otherwise.

For the case of sparse or irregular functional data, a fully non-parametric approach denoted as PACE \cite{yao2005} is used. Let $Y_{ij^{(h)}}$ be the $j$-hth observation obtained from CFD (\ref{5}) evaluating the $i$-th curve at time $t_{ij^{(h)}}\in [a,T_i^*]$, where $h=t_{ij^{(h)}}-t_{i{j-1}^{(h)}}$ is the step size chosen for the time increment by considering $t_{i1}=t_{i1^{(h)}}\leq t_{i2^{(h)}}\leq ...\leq t_{iJ_{i-1}^{(h)}}\leq  t_{iJ_{i}^{(h)}}= t_{iJ_i} $ for $i=1,...,N$. Then, including additional measurement errors $\varepsilon_{ij}$, uncorrelated among them with $E[\varepsilon_{ij^{(h)}}]=0$ and $Var[\varepsilon_{ij^{(h)}}]=\sigma^2$, to a model with measurements $Y_{ij^{(h)}}$ calculated at time $t_{ij^{(h)}}$ from (\ref{7}) we obtain:
\begin{equation}
\label{8}
Y_{ij^{(h)}}=X_i(t_{ij^{(h)}})+\varepsilon_{ij^{(h)}}=\mu(t_{ij^{(h)}})+\sum_{m=1}^\infty\nu_{im}\xi_m(t_{ij^{(h)}})+\varepsilon_{ij^{(h)}}\quad\text{with}\quad t_{ij^{(h)}}\in \mathcal{T}\,,
\end{equation} 
where the expansion (\ref{8}) can be approximated as follows:
\begin{equation}
\label{9}
Y_{ij^{(h)}}\approx\mu(t_{ij^{(h)}})+\sum_{m=1}^p \nu_{im}\xi_m(t_{ij^{(h)}})+\varepsilon_{ij^{(h)}}\quad\text{with}\quad t_{ij^{(h)}}\in \mathcal{T}.
\end{equation}

Moreover, the mean function, the covariance function and the eigenfunctions are calculated using the local linear smoothers \cite{fan1992variable}. From (\ref{8}), we note that the $Cov(Y_{ij^{(h)}},Y_{il^{(h)}}|\,t_{ij^{(h)}},t_{il^{(h)}}=Cov(X(t_{ij^{(h)}},X(t_{il^{(h)}})+\sigma^2\delta_{jl}$ for which $G_i(t_{ij^{(h)}},t_{il^{(h)}})=(Y_{ij^{(h)}}-\hat{\mu}(t_{ij^{(h)}})\,(Y_{il^{(h)}}-\hat{\mu}(t_{il^{(h)}})$, where $\hat{\mu}(t)$ is a mean function obtained by applying a local linear smoother to the scatterplot $\{(t_{ij^{(h)}},Y_{ij^{(h)}}): 1\leq i\leq N,\, 1\leq j \leq J_{i}^{(h)}\}$. The estimated covariance surface, $\hat{G}(s,t)$, is calculated considering the raw estimates $G_i(t_{ij^{(h)}},t_{il^{(h)}})$ and then applying two-dimensional smoothing to the scatterplot 
$\{(\,(t_{ij^{(h)}},t_{il^{(h)}})\,;\, G_i(t_{ij^{(h)}},t_{il^{(h)}})\,): 1\leq i\leq N,\, 1\leq j\neq l \leq J_{i}^{(h)}\}$.
Nevertheless, the calculation of scores $\nu_{im}$ is done according to the definition of integrals in $L^2(\mathcal{T})$, but in this case having that $Y_{ij^{(h)}}$ are available only at discrete random times $t_{ij^{(h)}}$ we can obtain the estimates $\hat{\nu}_{im}$ from (\ref{8}) as $\hat{\nu}_{im}=\sum_{j=1}^{J_i^{h}}(Y_{ij^{(h)}}-\hat{\mu}(t_{ij^{(h)}})\,\hat{\xi}_{m}(t_{ij^{(h)}})\,(t_{ij^{(h)}}-t_{ij-1^{(h)}})$. The goal is to obtain predicted trajectories $\hat{X}_i(t)$ for irregular data $Y_{ij^{(h)}}$. 
In summary, by discretising continuous trajectories with step sizes $h$, we can effectively analyse functional data even when observations are sparse or irregular. This approach allows for the practical implementation of FPCA, capturing essential features of the data while accounting for measurement errors and irregular observation patterns.

An alternative to the Riemann sum given by the previous formula is to assume that in (\ref{8}), the $\nu_{im}$ and $\varepsilon_{ij^{(h)}}$ are jointly Gaussian. For this reason, defining $\bm{\hat{Y}}_{i}^{(h)}=(Y_{i1^{(h)}},...,Y_{iJ_{i}^{(h)}})^T$, $\bm{\mu}_i^{(h)}=(\mu(t_{i1^{(h)}}),...,\mu(t_{iJ_{i}^{(h)}}))^T$ and $\bm{\xi}_{im}^{(h)}=(\xi_m(t_{i1^{(h)}}),...,\xi_m(t_{iJ_{i}^{(h)}}))^T$ under Gaussian assumptions, we obtain the best prediction for the $m$-th FPC score $\nu_{im}$ for the $i$-th subject as conditional expectation $E[\nu_{im}|\bm{\hat{Y}}_{i}^{(h)}]$ as follows:
\begin{equation}
\hat{\nu}_{im}=E[\nu_{im}|\bm{\hat{Y}}_{i}^{(h)}]=E[\nu_{im}]+Cov(\nu_{im},\bm{\hat{Y}}_{i}^{(h)})Cov(\bm{\hat{Y}}_{i}^{(h)},\bm{\hat{Y}}_{i}^{(h)})^{-1}(\bm{\hat{Y}}_{i}^{(h)}-\bm{\mu}_i^{(h)})=\lambda_m\bm{\xi}_{im}^{(h)\,T}\bm{\Sigma}_{\bm{\hat{Y}}_{i}^{(h)}}^{-1}(\bm{\hat{Y}}_{i}^{(h)}-\bm{\mu}_i^{(h)})\,,
\label{10}
\end{equation} 
where 
$\bm{\Sigma}_{\bm{\hat{Y}}_{i}^{(h)}} = \text{Cov}(\bm{\hat{Y}}_{i}^{(h)}, \bm{\hat{Y}}_{i}^{(h)}) = \text{Cov}(\bm{\hat{X}}_{i}^{(h)}, \bm{\hat{X}}_{i}^{(h)}) + \sigma^2 \bm{I}$, 
with $\bm{\hat{X}}_{i}^{(h)} = (\bm{X}(t_{i1^{(h)}}), ..., \bm{X}(t_{iJ_{i}^{(h)}}))$ and $\bm{I} \in \mathbb{R}^{J_{i}^{(h)} \times J_{i}^{(h)}}$ identity matrix. 
Thus, $(\bm{\Sigma}_{\bm{\hat{Y}}_{i}^{(h)}})_{jl} = G(t_{ij^{(h)}}, t_{il^{(h)}}) + \sigma^2 \delta_{jl}$, where $\delta_{jl} = 1$ if $j = l$ and $0$ otherwise. 
Equation (\ref{10}) can be obtained by using the estimates of $\lambda_m$, $\bm{\xi}_{im}^{(h)}$, and $\bm{\Sigma}_{\bm{\hat{Y}}_{i}^{(h)}}$ as follows:

\begin{equation}
\hat{\nu}_{im}=\hat{\lambda}_m\hat{\bm{\xi}}_{im}^{\,(h)\,T}\hat{\bm{\Sigma}}_{\bm{\hat{Y}}_{i}^{(h)}}^{-1}(\bm{\hat{Y}}_{i}^{(h)}-\hat{\bm{\mu}}_i^{(h)}).
\label{11}
\end{equation}
Finally, from (\ref{11}) the prediction for the trajectory $X_i(t)$ for $i$th subject can be done considering the first $p$ eigenfunctions as follows:
\begin{equation}
\label{12}
\hat{X}_i(t)=\hat{\mu}(t)+\sum_{m=1}^p\hat{\nu}_{im}\hat{\xi}_m(t)\quad\text{with}\quad t\in \mathcal{T}.
\end{equation}

\subsubsection{Functional Survival Tree (FST)}

In supervised learning, Survival Trees (STs) extend the classical Classification Trees (CTs) grown on censored data. Unlike classical decision trees, STs are not directly applicable to censored data; a ST focuses on predicting the time until an event of interest occurs for a subject. These problems introduce a distinctive challenge when dealing with survival data, making certain aspects of implementing ST more intricate than decision trees for classification tasks. In survival studies, we have, for each subject, an observation vector $\bm{Y}_i=(Y_{i1},..., Y_{iJ_{i}})^T$ that is a data collected in $J_i$ values until the event time $T_i^*=\mbox{min}\,(C_i, T_i)$ occurs for the $i$-th unit and the response vector $\bm{Z}=(Z_1, Z_2,..., Z_N)^T$ that can have binary values given by $Z_i=0$ or $Z_i=1$ for $i=1,..., N$. Thus, we observe, for the $i$-th subject, the data given by $(\bm{Y}_i,\bm{t}_i,Z_i,T_i^*,\delta_i)$, with $\delta_i$ censoring indicator.

In traditional CTs, the non-pruned trees lead to identifying homogeneous terminal nodes (leaves) using some splitting criteria based on decreasing an impurity metric, e.g. the Gini or Shannon-Weier indexes. However, in STs, the splitting criterion used in each node $s$ of the tree, denoted by $T$, differs from classical methods when dealing with censored data. Indeed, iteratively, the procedure grows a tree until each node defines, at the very least, a singular distinct event by using survival splitting as a means for maximizing between-node survival differences given by curves defined through parametric (Cox regression) and non-parametric (Kaplan-Meier) approaches \cite{Shimokawa2015ComparisonOS}. 

In a classical decision tree, the features vectors $\bm{X}\in\mathbb{R}^J$ have the same size for each unit: on the contrary, in STs the challenge is how to grow a tree having, for each subject, data collected during follow-up study $\mathcal{T}$ given by $Y_{ij}=Y(t_{ij})$ for $j=1,..., J_i$ values with $J_i\neq J_k$ for some $i\neq k$ with $t_{ij}\in \mathcal{T}$. For this reason, starting from these data types and defining the CFD, we deal with new features using the values estimated on CFD and applying the PACE decomposition on them. 
Therefore, extending Functional Classification Trees \cite{MaturoVerde} to the context of survival analysis, we propose the Functional Survival Classification Tree with Principal Components (FSCTs-FPCs). The latter classifier considers the new data $(\bm{\hat{\nu}}_i, Z_i, T_i^*,\delta_i)$, where $\bm{\hat{\nu}}_i=(\hat{\nu}_{i1},\hat{\nu}_{i2},...,\hat{\nu}_ {ip})^T$ corresponds to the $i$-th score vector obtained from Equation \ref{11}. Hence, the features matrix is defined as follows:

\begin{equation}
\bm{\hat{V}}=
\begin{pmatrix}
\hat{\nu}_{11} & \hat{\nu}_{12}  & \dots & \hat{\nu}_{1p}\\
\hat{\nu}_{21}  & \hat{\nu}_{22}  & \dots & \hat{\nu}_{2p} \\
\vdots & \vdots & \ddots & \vdots \\
\hat{\nu}_{N1}  & \hat{\nu}_{N2}  & \dots & \hat{\nu}_{Np} \\
\end{pmatrix}
\end{equation}

\noindent where $\bm{\hat{V}}\in \mathbb{R}^{N\times p}$ and $\hat{\nu}_{im}$ is the $m$-th FPC score of the $i$-th curve relative to the eigenfunction $\hat{\xi}_m$ for $i=1,...,N$ and $m=1,...,p$.

Typically, considering the node $s$ of the survival tree $T$, given a score-predictor $\bm{\hat{\nu}}_m$, the method finds a value $c$ and selects the best $\bm{\hat{\nu}}_0$ such that survival differences between the two inequalities, $\bm{\hat{\nu}}_0\leq c$ and $\bm{\hat{\nu}}_0 >c$, are maximized in order to predict the response $\bm{Z}$. The aim is to define in node $s$ two regions $R_1(m,c)=\{\hat{X}(t)\in L^2(\mathcal{T}) | \bm{\hat{\nu}}_m\leq c\}$ and $R_2(m,c)=\{\hat{X}(t)\in L^2(\mathcal{T}) | \bm{\hat{\nu}}_m > c\}$ to identify $l=1,...,L$ times for the two groups, with $t_1\leq t_2\leq ... \leq t_L$. One of the survival criteria used within node $s$, to select the threshold $c$ and the functional score-prediction $\bm{\hat{\nu}}_0$, is the log-rank test as splitting method \cite{ziegler2007}. It can be used to
maximize a splitting value $c$ for a predictor variable $\bm{\hat{\nu}}$ as follows:
\begin{equation}
\label{13}
|L(\bm{\hat{\nu}},c)|=\frac{\sum_{l=1}^{L} \Big(d_{l1}-r_{l1}\frac{d_{l}}{r_{l}}\Big)}{\sqrt{\sum_{l=1}^{L}\frac{r_{l1}}{r_{l}}\Big(1-\frac{r_{l1}}{r_{l}}\Big)\Big(\frac{r_{l}-d_{l}}{r_{l}-1}\Big)d_{l} } } 
\end{equation} 
with $d_{lj}$ number of events in daughter nodes $j\in\{1,2\}$ at time $t_{l}\in \mathcal{T}$ (time of observation), where $d_{l}=d_{l1}+d_{l2}$ is the total number of events within node $s$. Moreover, $r_{lj}$ is the number of individuals who are at risk (alive) in daughter nodes $j\in\{1,2\}$ at time $t_{l}\in\mathcal{T}$, where $r_{l}=r_{l1}+r_{2}$ is the total number of individuals alive within node $s$. $K$ is the total number of times between two daughter nodes. The number $r_{l1}$ corresponds to $T_{l}\ge t_{l}$ for which $\bm{\hat{\nu}}\le c$, while $r_{l2}$ corresponds to $T_{l}\ge t_{l}$ for which $\bm{\hat{\nu}} > c$ within node $s$. The Equation (\ref{13}) allows finding the best $\bm{\hat{\nu}}^*$ and the best $c^{*}$ which $|L(\bm{\hat{\nu}}^{*},c^{*})|\geq|L(\bm{\hat{\nu}},c)|$ for each $\bm{\hat{\nu}}$ and $c$ chosen randomly. This process is repeated randomly at every node $s$ until the terminal node is reached. The FSCTs-FPCs use different methods for constructing predictive models. Two common alternative approaches include log-rank splitting, which divides data based on survival differences, and the Nelson-Aalen estimator, used for estimating cumulative hazard. The last method helps us to identify significant predictors of survival outcomes in the terminal nodes. Once the survival tree $T$ is built, let $s$ be the terminal node, where $s=1,...,S$, in which are defined the times $t_{1s}\leq t_{2s}\leq ... \leq t_{ L_{s}s} $ and let $d_{ls}$ and $r_{ls}$ the number of events and individuals at risk at time $t_{ls}$, the survival predictor in $T$ is defined in terms of the predictor within each terminal node $s$. For this, the Cumulative Hazard Function (CHF) and the Survival Function (SF) are obtained by using the Nelson-Aalen estimator and Kaplan-Meier estimator defined as follows:
\begin{equation}
\mathcal{\hat{H}}_s(t) = \sum_{t_{ls} \leq t} \frac{d_{ls}}{r_{ls}} , \quad \hat{S}_s(t) = \prod_{t_{ls} \leq t} \Bigg(1 -\frac{d_{ls}}{r_{ls}}\Bigg)\,
\label{chf}
\end{equation}
where $\mathcal{H}_s(t)$ and $\hat{S}_s(t)$ are the hazard and the survival estimate for a terminal node $s$ obtained considering all distinct event times $t_{ls} \leq t$.
However, predicting survival outcomes after the survival tree construction is essential. This evaluation often involves comparing the predicted survival probabilities with the observed outcomes. One standard method for evaluating the importance of a survival tree is using the Concordance Index (C-index), also known as Harrell's C or the C-statistics, which tells us how accurately the model ranks people's survival times. A higher C-index indicates better predictive performance of the survival tree model. Other metrics, such as the Brier score, can be used for further evaluation.

\subsubsection{Functional Random Survival Forest (FRSF)} 

The Functional Random Survival Forest (FRSF) extends the classic RSF. In our approach, the method is built for the CFD considering the pairs $\{Z_i, X_i(t)\}$, where $X_i(t)$ is a CFD curve for $i$-th unit defined in the metric space $(L^2(\mathcal{T}),\Vert\cdot\Vert_2)$, while $Z_i$ is a scalar response. 
Starting from the dataset $D=\{Z_i, \hat{\bm{\nu}}_i, T_i^*,\delta_i\}_{i=1}^N$ used for the realization of the model, where $\hat{\bm{\nu}}_i\in \mathbb{R}^{p}$ is the $i$-th score calculated by PACE on CFD, $Z_i$ is a binary response, $T_i^*=\min\,(C_i, T_i)$, and $\delta_i$ is a censoring indicator at sample $i=1,..., N$. The method generates $B$ bootstrap samples from $D$ by creating the Training data set (In-bag data) and Validation data set (Out-of-bag data) to construct multiple STs. 

The following pseudo-code can summarise the FRSF: 

\begin{algorithm}
	\caption{Functional Random Survival Forest (FRSF)}
	\begin{algorithmic}[1]
    \State \textbf{Input:} Dataset $D = \{(Z_i,\hat{\bm{\nu}}_i, T_i^*, \delta_i)\}_{i=1}^N$, number of trees $B$, number of random predictors $q$
		\For {$b = 1,2,...,B$}
        \State Sample a $b$ bootstrap sample of size $N$ from  $D$ with replacement (In-bag data)
        \State Use the remaining data as Out-of-bag data
        \State Train a survival tree $T_b$ on the bootstrap sample $b$ using $q$ random FPCs' scores at the node
        \State Calculate a CHF in each terminal node $s$
		\EndFor
  \State Calculate the average  CHF over all survival trees $T_1,T_2,\ldots,T_B$ for IB and OOB data
  \State \textbf{Output:} Survival Trees $\{T_1, T_2, ..., T_B\}$ and the averaged CHF 
	\end{algorithmic} 
 
\end{algorithm}

The step $8$ of the algorithm can be explained from (\ref{chf}) as follows: 

\begin{equation}
\overline{\mathcal{H}}^{IB}(t|\hat{\bm{\nu}})=\frac{1}{B}\sum_{b=1}^{B} {\mathcal{H}}_b(t|\hat{\bm{\nu}})\,,\qquad\overline{\mathcal{H}}_i^{OOB}(t|\hat{\bm{\nu}}_i)=\frac{1}{B}\sum_{b=1}^{B} {\mathcal{H}}_b(t|\hat{\bm{\nu}}_i)
\end{equation}
where ${\mathcal{H}}_b(t|\hat{\bm{\nu}})$ is the in-bag CHF, while ${\mathcal{H}}_b(t|\hat{\bm{\nu}}_i)$ is the out-of-bag CHF for $b$-th survival tree. 

The metrics derived from the ensemble CHF, such as the predictor error and the variable importance (VIMP), provide crucial insights into the model's goodness-of-fit calculated using only the OOB data. This process evaluates the accuracy of predictors and determines their relative importance in predicting survival outcomes. In FRSF the Breiman-Cutler Variable Importance (VIMP), also known as permutation importance \cite{breiman2002manual} is used to quantify the importance of each predictor variable by evaluating its impact on prediction error. The VIMP method permutes the out-of-bag (OOB) values of $\hat{\bm{\nu}}$ within each tree. This permutation process entails randomising the values of 
$\hat{\bm{\nu}}$, while keeping the values of the other predictor variables unchanged. Moreover, the modified data influences decision paths through the tree structure based on the altered $\hat{\bm{\nu}}$ values. The resulting out-of-bag error, indicative of prediction accuracy with the permuted $\hat{\bm{\nu}}$ is compared to the original error calculated with unaltered $\hat{\bm{\nu}}$ values. The discrepancy between these errors quantifies the importance of 
$\hat{\bm{\nu}}$ in predicting outcomes. Aggregating these individual importance scores across all trees in the forest yields the permutation importance for variable $\hat{\bm{\nu}}$, providing insights into its contribution to the overall predictive performance of the model.

\section{Application}

\begin{figure}[htbp!]
  \centering
  \includegraphics[width=0.8\linewidth]{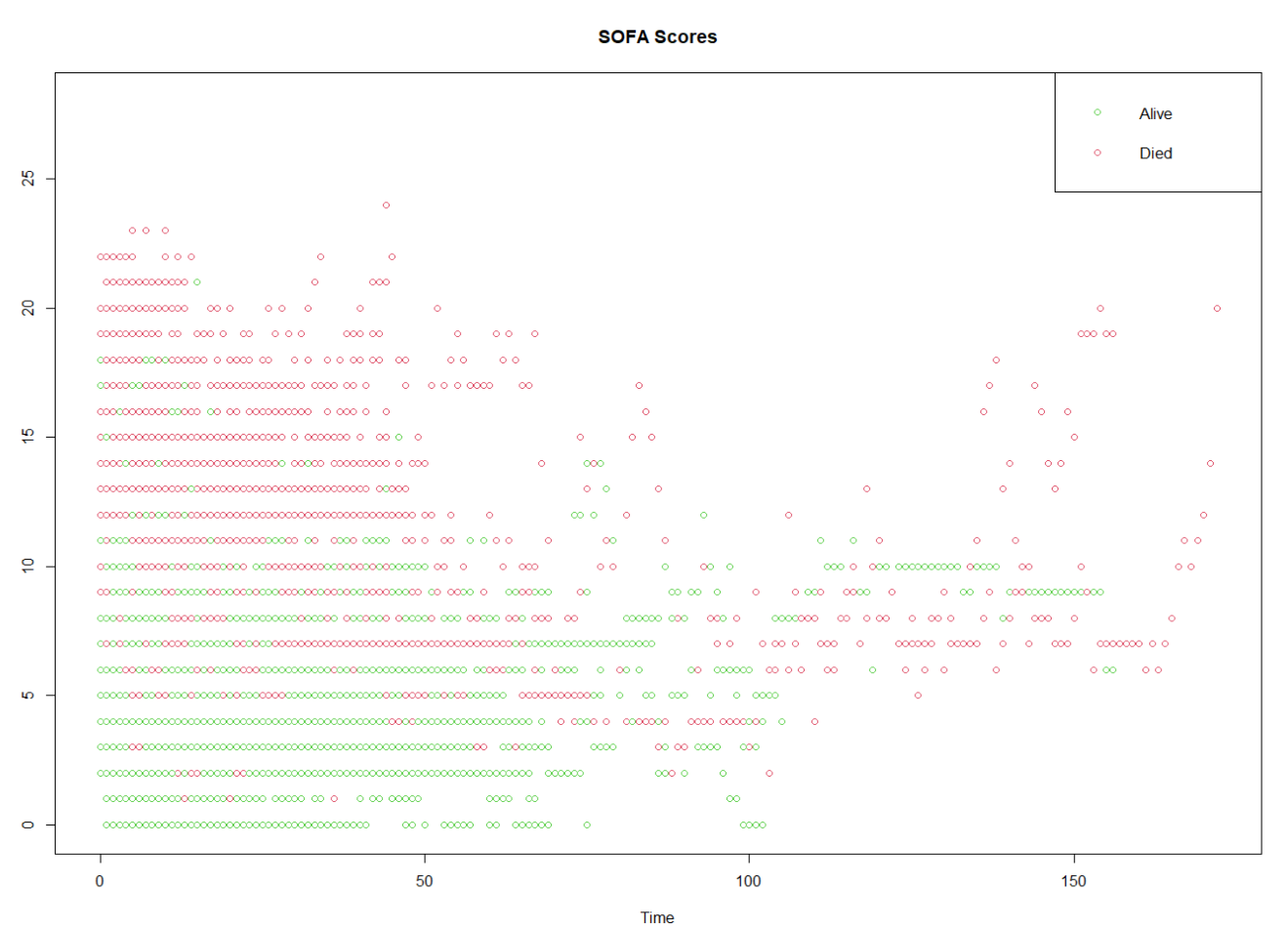}
  \caption{SOFA scores collected in $173$ ICU days concerning outcome}
  \label{fig:image1}
\end{figure}

\begin{figure}[htbp!]
  \centering
  \includegraphics[width=\linewidth]{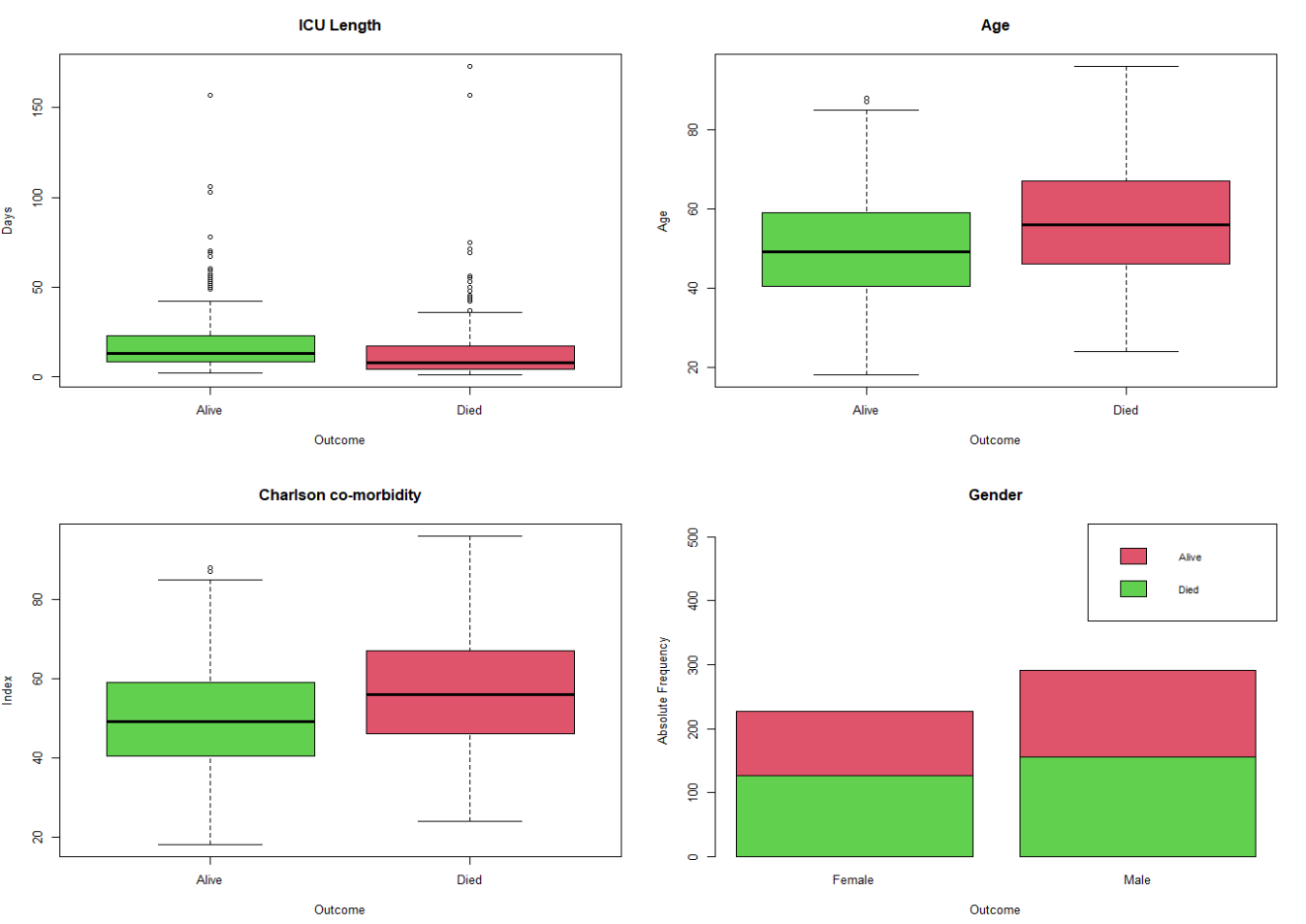}
  \caption{Boxplots and Barplot data for the variables concerning outcome}
  \label{fig:image2}
\end{figure}

We consider a dataset from the field of critical care medicine called the Sequential Organ Failure Assessment (SOFA) score, designed to provide a comprehensive evaluation of organ function in critically ill patients. The SOFA score dataset is available in the R package 'refund'. Many applications have been developed utilizing the SOFA score dataset, as evidenced by several studies \cite{moreno2023}. This dataset concerns specifically $520$ patients admitted to the Intensive Care Unit (ICU) with Acute Lung Injury. For each patient, daily measurements of SOFA scores, which range from $0$ to $24$, have been recorded for each patient over $173$ ICU days as shown in Figure (\ref{fig:image1}), indicating the severity of organ failure. 
Other variables collected during hospitalization include ICU death indicators, ICU length of stay (LOS), patient age, gender, and the Charlson co-morbidity index, providing insights into baseline health status. Figure (\ref{fig:image2}) shows descriptive statistics for the variables.
Boxplots visually summarize the distributional characteristics of the numerical variables, while barplots compare categorical data. All data concerning the outcome variable indicate the survival status, corresponding to $237$ patients died and $283$ lived until the last recording time.

\begin{figure}[htbp!]
    \centering
    \includegraphics[width=0.8\linewidth]{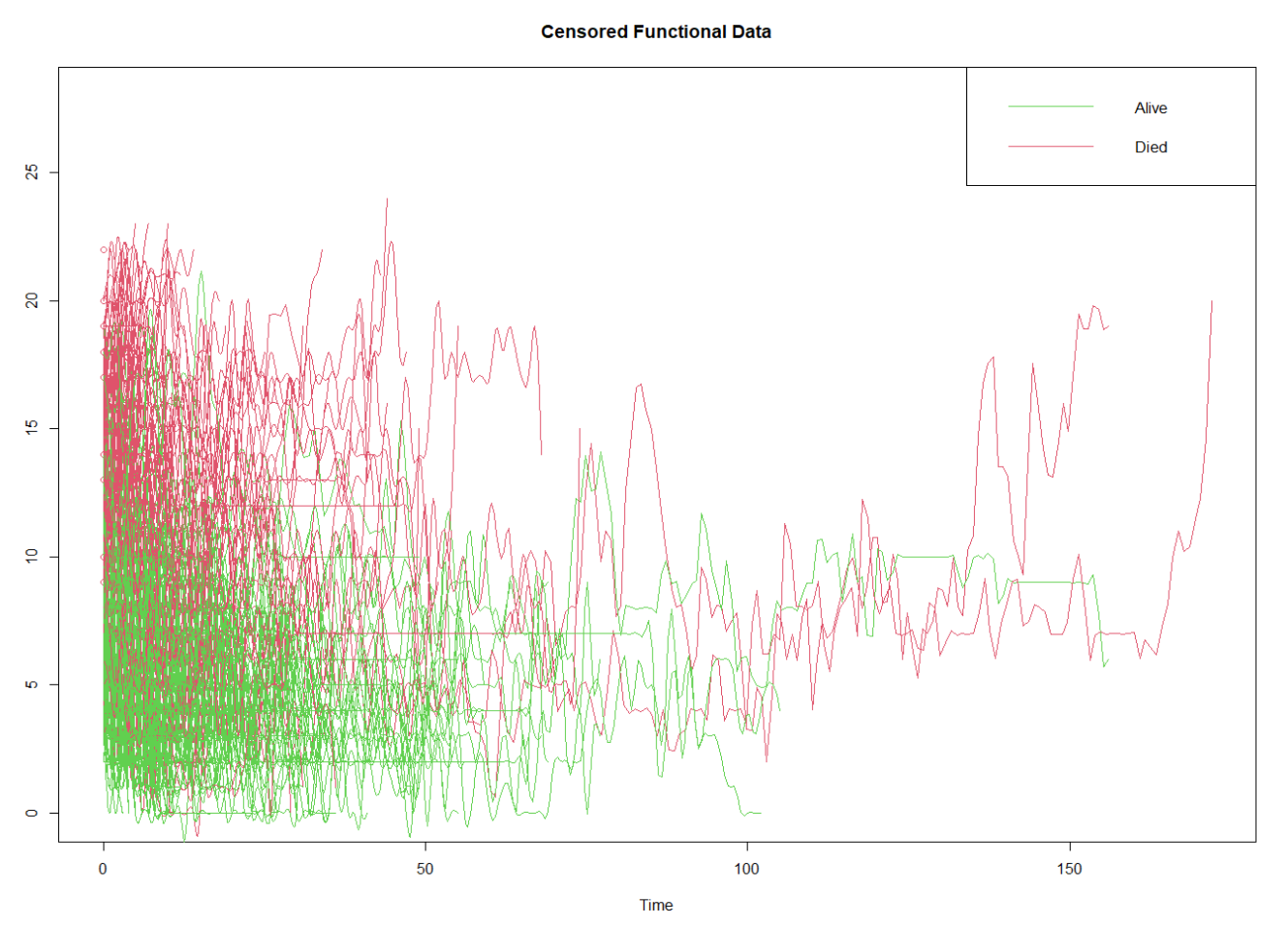}
    \caption{CFD for subjects concerning outcome}
    \label{fig3}
\end{figure}

The first problem we address is converting the SOFA scores into a functional form to monitor recovery progress throughout hospitalization. We thus propose a functional approach based on the previously described definition in Equation \ref{CFD}, where the response variable corresponds to the SOFA scores, and the independent variable represents the $ t$ at which the score was calculated for different recorded values. For each statistical unit, truncated curves will be defined based on the scores recorded during hospitalization as follows:

    \begin{align}
        \label{eq:case1}
        \hat{Y}_{\text{score}}(t)_{(1)} &= Y_{\text{score}} \\
        \label{eq:case2}
        \hat{Y}_{\text{score}}(t)_{(2)} &= \hat{\beta}_0+\hat{\beta}_1 t  \\
        \label{eq:case3}
        \hat{Y}_{\text{score}}(t)_{(i)} &= \hat{c}_{1}\hat{\phi_1}(t)+\hat{c}_{2}\hat{\phi_2}(t)+\ldots+ \hat{c}_{K_i}\hat{\phi}_{K_i}(t) & \text{with}\quad & i \in \Omega & \Omega=\{i \in \mathcal{I}: J_i\geq 3\}
    \end{align}

\noindent where in the equation (\ref{eq:case3}) we select the set of functions $\{\hat{\phi}_k\}_{k=1,\ldots,K_i}$ as B-splines of order $4$. 
The truncated curves have been visualized respect to the corresponding outcome variable, as shown in the Figure (\ref{fig3}). To assess the model quality of the FRSF, we propose to use different training datasets, comprising $50\%$, $60\%$, $70\%$, and $80\%$ of the SOFA dataset. 

\begin{table}[h!]
\centering
\begin{tabular}{c|c|c|c|c|c|c}
\hline
\multirow{2}{*}{Train Dataset} & \multicolumn{2}{c|}{STD Model} & \multicolumn{2}{c|}{CFD Model(h=0.5) } & \multicolumn{2}{c}{ CFD Model(h=0.2)} \\ \cline{2-7} 
                       &(OOB) CRPS & (OOB) RPE & (OOB) CRPS & (OOB) RPE & (OOB) CRPS & (OOB) RPE\\ \hline
50\%                 & 0.11830311 & 0.14836430 & 0.11861220  & 0.14734700 & 0.11130006 & 0.13733469\\ \hline
60\%                 & 0.11631665 & 0.14853905 & 0.11517514 & 0.14402103 & 0.11273948 & 0.14061401 \\ \hline
70\%                  & 0.09189498 & 0.14906593& 0.09253855  & 0.14428571 & 0.08753912 &  0.14195055 \\ \hline
80\%                   & 0.10149814 & 0.14921755 & 0.10135369 & 0.14619833 & 0.09974069  & 0.14570930\\ \hline
\end{tabular}
\caption{Comparison of OOB scores for CRPS and RPE across different training datasets.}
\label{quattro}
\end{table}

The model is evaluated considering the four scenarios obtained from the previously defined partition. 
In particular, we compare our proposed method (FRSF for CFD) and the traditional FRSF for STD data. 
We assess the models' performance across the different training dataset proportions using various metrics, such as the Continuous Ranked Probability Score (CRPS) and the Requested Performance Error (RPE) on the OOB data proportion. 
As seen in Table \ref{quattro}, the models show varying degrees of accuracy depending on the proportion of the training dataset. 

A more detailed analysis of temporal evaluations with different temporal discretisation values, represented by the parameter $h$, enables capturing information that may otherwise be lost during the follow-up period. By carefully selecting $h$, we ensure a more detailed and accurate temporal assessment, filling in gaps where the information was not collected initially. Specifically, we have chosen two different bandwidth values, $h=0.5$ and $h=0.2$, within the CFD model. 
Moreover, the model's performance has been analyzed using the RPE scores based on the number of trees and the CRPS scores to observe the forecasted temporal trend. Graphs for the four different scenarios (\ref{four_figures2}) illustrate the model's predictive performance evolution.
\begin{figure}[htbp!]
    \centering
    \includegraphics[width=0.82\textwidth]{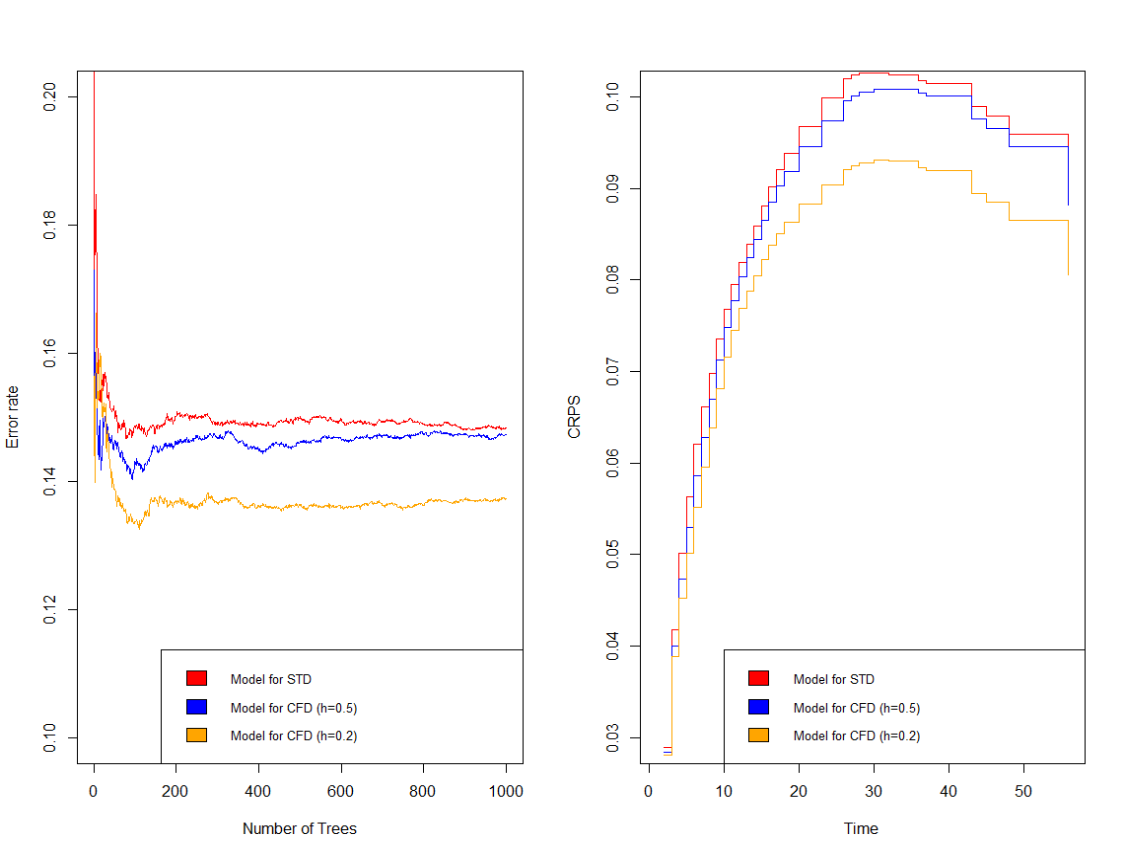}
    \caption*{(a) Case 50\%}
    \label{fig:fig1}
    
    \vspace{1em} 

    \includegraphics[width=0.82\textwidth]{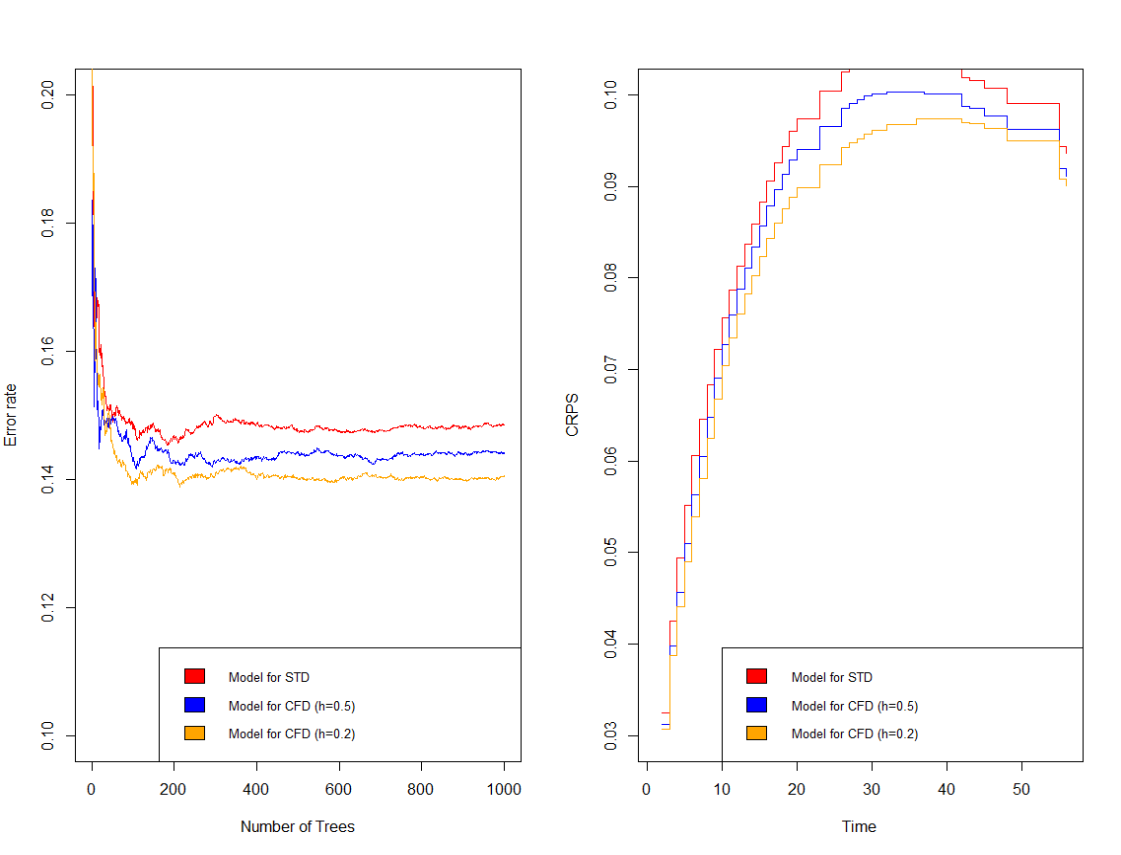}
    \caption*{(b) Case 60\%}
\end{figure}

    \begin{figure}
    \centering
        \includegraphics[width=0.82\textwidth]{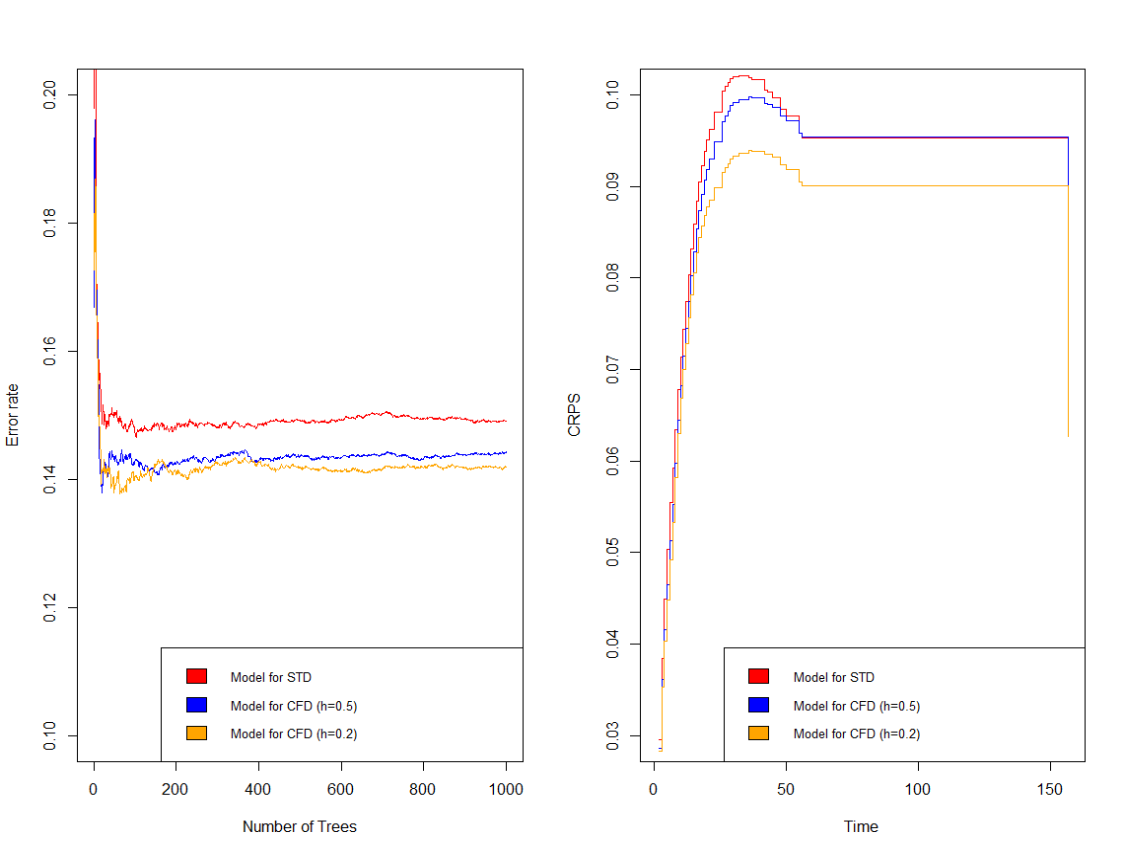}
        \caption*{(c) Case 70\%}
        \label{fig:fig3}

    \vspace{1em} 
        \includegraphics[width=0.82\textwidth]{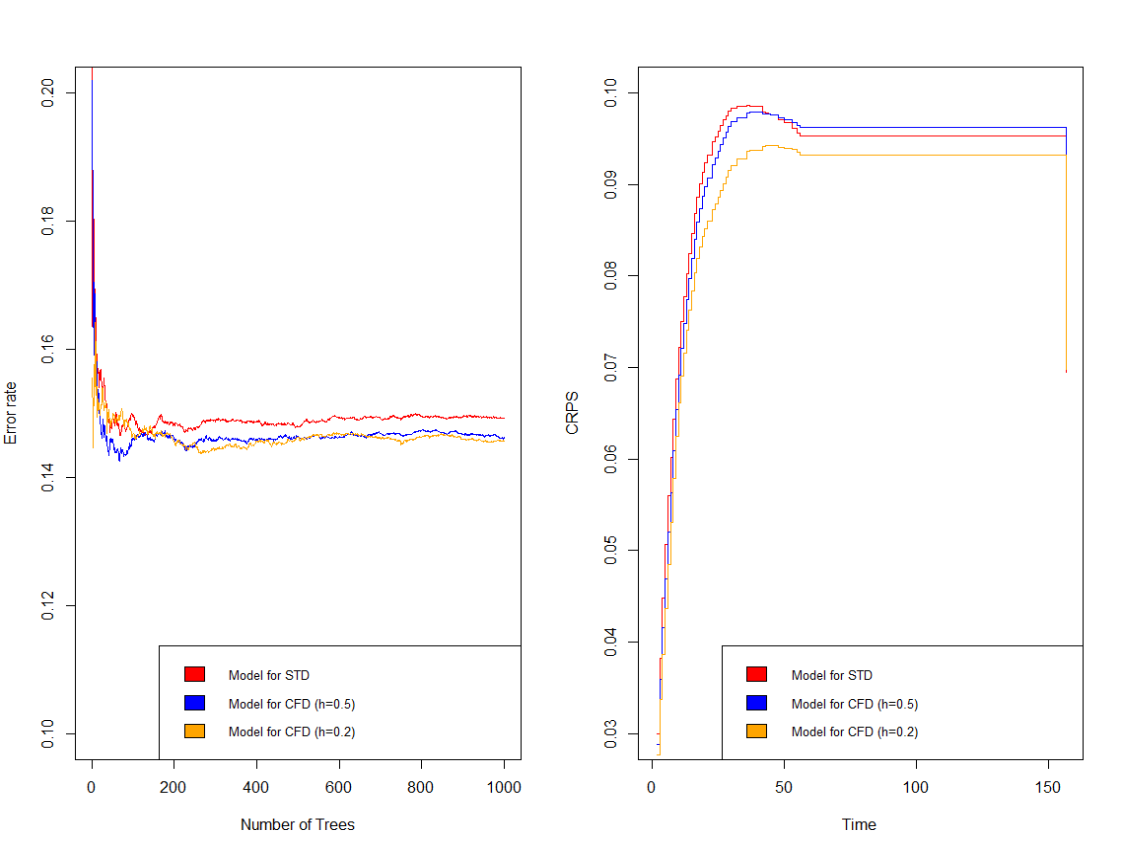}
        \caption*{(d) Case 80\%}
        \label{fig:fig4}
    \caption{Error rate and CRPS function calculated for all scenarios}
    \label{four_figures2}
\end{figure}
It can be noted that there is a significant improvement in the RPE (left panel) and CRPS (right panel) in all cases. 

The CRPS function shows better performance for the FRSF with CFD during intermediate daily periods when a substantial amount of data is available.
For example, when examining the CRPS and RPE graphs for the Model CFD with $h=0.2$, we observe a gradual decrease in both metrics over time, indicating improved model accuracy.
Furthermore, the quality of the models has been shown through Variable Importance (VIMP) and Relative Importance (RI) for scenarios (50\%, 60\%, 70\%, and 80\% of the Training dataset) for the three models. In our specific analysis, we focus on the scenario using 80\% of the training dataset (results for other scenarios are provided in the appendix). The variables considered include the first four FPCs scores, Age, Charlson Comorbidity Index, and Gender. By examining the variable importance values across different scenarios and models, we can determine each feature relative contribution to the predictive performance of the models (Table \ref{table:VIMP}). 

\begin{table}[htbp]
\centering
\begin{minipage}{0.4\textwidth}
\centering
\begin{tabular}{c|c|c}
\hline
 Variable & Importance & Relative Importance \\ 
 \hline
    PC1      & 0.4138     & 1.0000       \\
    PC2      & 0.1325     & 0.3201       \\
    PC4      & 0.0786     & 0.1900       \\
    PC3      & 0.0510     & 0.1231       \\
    Age      & 0.0435     & 0.1052       \\
    Charlson & 0.0284     & 0.0686       \\
    Gender   & -0.0004    & -0.0009      \\
\hline
\end{tabular}
\end{minipage}\hfil
\begin{minipage}{0.4\textwidth}
\centering
\begin{tabular}{c|c|c}
\hline
Variable & Importance & Relative Importance \\
\hline
    PC1      & 0.4202     & 1.0000       \\
    PC2      & 0.1287     & 0.3062       \\
    PC4      & 0.0806     & 0.1919       \\
    PC3      & 0.0487     & 0.1159       \\
    Age      & 0.0416     & 0.0989       \\
    Charlson & 0.0270     & 0.0644       \\
    Gender   & -0.0008    & -0.0019      \\
\hline
\end{tabular}
\end{minipage}

\vspace{1cm}

\begin{minipage}{0.4\textwidth}
\centering
\begin{tabular}{c|c|c}
\hline
Variable & Importance & Relative Importance \\
\hline
PC1      & 0.4396     & 1.0000       \\
    PC2      & 0.1191     & 0.2709       \\
    PC4      & 0.0945     & 0.2149       \\
    PC3      & 0.0532     & 0.1211       \\
    Age      & 0.0434     & 0.0987       \\
    Charlson & 0.0346     & 0.0786       \\
    Gender   & -0.0004    & -0.0008      \\ 
\hline
\end{tabular}
\end{minipage}
\caption{Variable Importance (VIMP) for Model employing 80\% of the data as the training set. The top-left table corresponds to the Model STD, the top-right table corresponds to the Model CFD (h=0.5), and the bottom table represents the Model CFD (h=0.2).
}
\label{table:VIMP}
\end{table}
In general, PC1 consistently exhibits the highest importance across all cases and models, followed by PC2 and PC4. Age and Charlson Comorbidity Index also show moderate importance, while Gender has an insignificant impact on the model. These values underscore the varying degrees of influence that different variables have on the predictive accuracy of the models.

\section{Discussion and Conclusions}

This study presents a new approach to survival analysis by developing Functional Random Survival Forest (FRSF), specially designed for Censored Functional Data (CFD). This approach embeds Functional Data Analysis within the framework of Random Survival Forest, thereby utilizing the strength of FPCA in managing complex, high-dimensional, censored, and temporally correlated data.

This study's critical contribution is introducing the FRSF model, a novel extension of traditional RSF incorporating functional data. This offers significant advantages compared to conventional methods, especially when the data are censored and/or high-dimensional. The implementation of CFD provides a robust framework for handling incomplete or irregularly spaced data by enabling the reconstruction of continuous trajectories from discrete observations and performing a more comprehensive analysis of survival dynamics. Following the work of Maturo and Verde on FRF \cite{Maturo2023}, the current study adapts and extends FRF to the context of survival analysis and mainly deals with the difficulties posed by censored data. Fusing FPCA into the FRSF framework allows for dimension reduction while retaining key features, which gives the model interpretability to the FST and explainability for the overall RSF model. That means the model remains flexible in accounting for differing observation times and censoring mechanisms, leading to robust inferences without some common assumptions made for the parametric survival models.

A crucial aspect of this study is the introduction of the parameter $h$, which denotes the step size for time increments in the evaluation of functional data. The parameter $h$ is meaningful as it permits discretizing continuous trajectories, enabling effective functional data analysis even when observations are sparse or irregular. Different values of $h$ can capture various levels of detail in the data, with smaller values providing finer resolution and potentially revealing more fine patterns in the temporal evolution of the data. Our analysis used different values of $h$  to demonstrate how different discretization levels impact the model's performance. The proposed FRSF model was empirically validated using the SOFA dataset, which consists of daily measurements of critically ill patients. 

Despite its strengths, the FRSF model has certain limitations. One of the primary challenges is the computational complexity associated with the FPCA decomposition and the subsequent construction of survival trees. The choice of basis functions and FPCs can also impact the model's performance. Future research should optimise these aspects to enhance the model's efficiency and scalability and test different basis functions, such as wavelets, to capture more complex and localised patterns in the data. 
In addition, future research could explore integrating FRSF with other machine learning approaches, such as boosting or deep learning, which could further enhance its predictive capabilities and scalability. 

In conclusion, the FRSF offers a powerful and versatile tool for survival analysis, particularly in the presence of censored and irregularly spaced functional data. Its integration of FDA and RSF techniques, along with empirical validation, demonstrates its potential for significant contributions to theoretical advancements and practical applications in survival analysis.

\subsection*{Funding and/or Conflicts of interests/Competing interests}

All the authors declare that they did not receive support from any organisation for the submitted work.
All authors certify that they have no affiliations with or involvement in any organization or entity with any financial or non-financial interest in the subject matter or materials discussed in this manuscript.

\bibliographystyle{abbrvnat}
\bibliography{bb.bib}%

\newpage
\FloatBarrier
\section*{Appendix}
\subsection*{VIMP Tables}
\begin{table}[!htbp]
\centering
\begin{minipage}{0.4\textwidth}
\centering
\begin{tabular}{c|c|c}
\hline
 Variable & Importance & Relative Importance \\ 
 \hline
   PC1     &       0.4466    &     1.0000\\
PC2      &      0.1209  &       0.2708\\
PC4        &    0.0610    &     0.1366\\
Age         &   0.0526   &      0.1177\\
PC3          &  0.0512  &       0.1147\\
Charlson      & 0.0398 &        0.0891\\
Gender        &-0.0004&        -0.0008\\
\hline
\end{tabular}
\end{minipage}\hfil
\begin{minipage}{0.4\textwidth}
\centering
\begin{tabular}{c|c|c}
\hline
Variable & Importance & Relative Importance \\
\hline
 PC1           & 0.4549  &       1.0000\\
PC2            & 0.1233  &      0.2710\\
PC4            & 0.0701  &     0.1541\\
Age            & 0.0497  &       0.1093\\
PC3            & 0.0459  &       0.1010\\
Charlson       & 0.0388  &       0.0854\\
Gender         & 0.0000  &       0.0000  \\
\hline
\end{tabular}

\end{minipage}

\vspace{1cm}
\begin{minipage}{0.4\textwidth}
\centering
\begin{tabular}{c|c|c}
\hline
Variable & Importance & Relative Importance \\
\hline
PC1           & 0.4749   &      1.0000\\
PC2           & 0.1144   &      0.2409\\
PC4           & 0.0800   &      0.1685\\
Age           & 0.0475   &      0.0999\\
PC3           & 0.0464   &      0.0977\\
Charlson      & 0.0381   &      0.0801\\
Gender        &-0.0001   &     -0.0002  \\ 
\hline
\end{tabular}
\end{minipage}
\caption{Variable Importance (VIMP) for 50\% of the training dataset. The top-left table corresponds to the Model STD, the top-right table corresponds to the Model CFD (h=0.5), and the bottom table represents the Model CFD (h=0.2).
}
\end{table}

\begin{table}[htbp]
\centering
\begin{minipage}{0.4\textwidth}
\centering
\begin{tabular}{c|c|c}
\hline
 Variable & Importance & Relative Importance \\ 
 \hline
PC1         &   0.4473   &      1.0000\\
PC2         &   0.1078   &      0.2409\\
PC4         &   0.0768   &      0.1716\\
Age         &   0.0656   &      0.1467\\
PC3         &   0.0467   &      0.1043\\
Charlson    &   0.0424   &      0.0947\\
Gender      &  -0.0007   &     -0.0015  \\
\hline
\end{tabular}
\end{minipage}\hfil
\begin{minipage}{0.4\textwidth}
\centering
\begin{tabular}{c|c|c}
\hline
Variable & Importance & Relative Importance \\
\hline
PC1        &    0.4571     &    1.0000\\
PC2        &    0.1117     &    0.2445\\
PC4        &    0.0824     &    0.1803\\
Age        &    0.0585     &    0.1280\\
PC3        &    0.0399     &    0.0873\\
Charlson   &    0.0389     &    0.0852\\
Gender     &   -0.0006     &   -0.0012\\
\hline
\end{tabular}
\end{minipage}

\vspace{1cm}

\begin{minipage}{0.4\textwidth}
\centering
\begin{tabular}{c|c|c}
\hline
Variable & Importance & Relative Importance \\
\hline
PC1      &      0.4674   &      1.0000\\
PC2      &      0.1094   &      0.2340\\
PC4      &      0.0916   &      0.1959\\
Age      &      0.0536   &      0.1147\\
PC3      &      0.0423   &      0.0904\\
Charlson &      0.0414   &      0.0886\\
Gender   &     -0.0003   &     -0.0006\\
\hline
\end{tabular}
\end{minipage}
\caption{Variable Importance (VIMP) for 60\% of the training dataset. The top-left table corresponds to the Model STD, the top-right table corresponds to the Model CFD (h=0.5), and the bottom table represents the Model CFD (h=0.2).
}
\end{table}

\begin{table}[htbp]
\centering
\begin{minipage}{0.4\textwidth}
\centering
\begin{tabular}{c|c|c}
\hline
 Variable & Importance & Relative Importance \\ 
 \hline
PC1     &       0.4338    &     1.0000\\
PC2     &       0.1169    &     0.2694\\
PC4     &       0.1067    &     0.2459\\
Age     &       0.0612    &     0.1411\\
PC3     &       0.0567    &     0.1308\\
Charlson&       0.0362    &     0.0836\\
Gender  &      -0.0001    &    -0.0003\\

\hline
\end{tabular}
\end{minipage}\hfil
\begin{minipage}{0.4\textwidth}
\centering
\begin{tabular}{c|c|c}
\hline
Variable & Importance & Relative Importance \\
\hline
PC1      &      0.4457    &     1.0000\\
PC2      &      0.1104    &     0.2477\\
PC4      &      0.1062    &     0.2382\\
Age      &      0.0557    &     0.1249\\
PC3      &      0.0515    &     0.1156\\
Charlson &      0.0432    &     0.0969\\
Gender   &     -0.0001    &    -0.0001\\
\hline
\end{tabular}
\end{minipage}

\vspace{1cm}

\begin{minipage}{0.4\textwidth}
\centering
\begin{tabular}{c|c|c}
\hline
Variable & Importance & Relative Importance \\
\hline
PC1        &    0.4465     &    1.0000\\
PC4        &    0.1045     &    0.2340\\
PC2        &    0.1024     &    0.2293\\
Age        &    0.0524     &    0.1174\\
PC3        &    0.0517     &    0.1157\\
Charlson   &    0.0415     &    0.0928\\
Gender     &   -0.0003     &   -0.0008\\
\hline
\end{tabular}
\end{minipage}
\caption{Variable Importance (VIMP) for 70\% of the training dataset. The top-left table corresponds to the Model STD, the top-right table corresponds to the Model CFD (h=0.5), and the bottom table represents the Model CFD (h=0.2).
}
\end{table}

\end{document}